\documentclass{ws-procs975x65}

\def\beq{\begin{equation}}
\def\eeq{\end{equation}}

\begin{document}

\title{Polarimetry and strong gravity effects from spots orbiting near a black hole}
\author{Vladim\'{\i}r Karas,\footnote{E-mail: vladimir.karas@cuni.cz} Michal Dov\v{c}iak, Ji\v{r}\'{\i} Svoboda, \& Wenda Zhang}
\address{Astronomical Institute, Czech Academy of Sciences, Bo\v{c}n\'{\i} II 1401, CZ-14100 Prague, Czech Republic}
\author{Giorgio Matt}
\address{Dip. Matematica e Fisica, Universit\`a Roma Tre, Via della Vasca Navale 84, I-00146, Rome, Italy}
\author{Andreas Eckart, \& Michal Zaja\v{c}ek\footnote{New address: Center for Theoretical Physics, Polish Academy of Sciences, Warsaw, Poland}} 
\address{I. Physikalisches Institut, Universit\"at zu K\"oln, Z\"ulpicher Str. 77, D-50937 Cologne, Germany\\
Max-Planck-Institut f\"ur Radioastronomie, Auf dem H\"ugel 69, D-53121 Bonn, Germany}

\begin{abstract}
We study the modulation of the observed radiation flux and the associated
changes in the polarization degree and angle that are predicted by the
orbiting spot model for flares from accreting black holes.
The geometric shape of the emission region influences the
resulting model lightcurves, namely, the emission
region of a spiral shape can be distinguished from a simpler
geometry of a small orbiting spot. 

We further explore this
scheme for the observed flares from the supermassive black hole in the
context of Galactic center (Sgr A*). Our code
simulates the lightcurves for a wide range of parameters.
The energy dependence of the changing degree and angle of polarization
should allow us to discriminate between the cases of a rotating and a
non-rotating black hole.
\end{abstract}
\bodymatter

\section{Introduction}
Relativistic corrections
to a signal from orbiting spots can lead to large rotation in the
plane of observed X-ray polarization. When integrated over an extended
surface of the source, this can diminish the observed degree of polarization. Such
effects are potentially observable and can be used to distinguish among
different models of the source geometry and the radiation mechanisms
responsible for the origin of the polarized signal.
The idea was originally proposed in the 1970s,\cite{con77,con80,pin77} however,
its observational confirmation and practical use in observations 
are still is a challenging task.

\begin{figure}[tbh!]
\begin{center}
\includegraphics[width=0.32\textwidth]{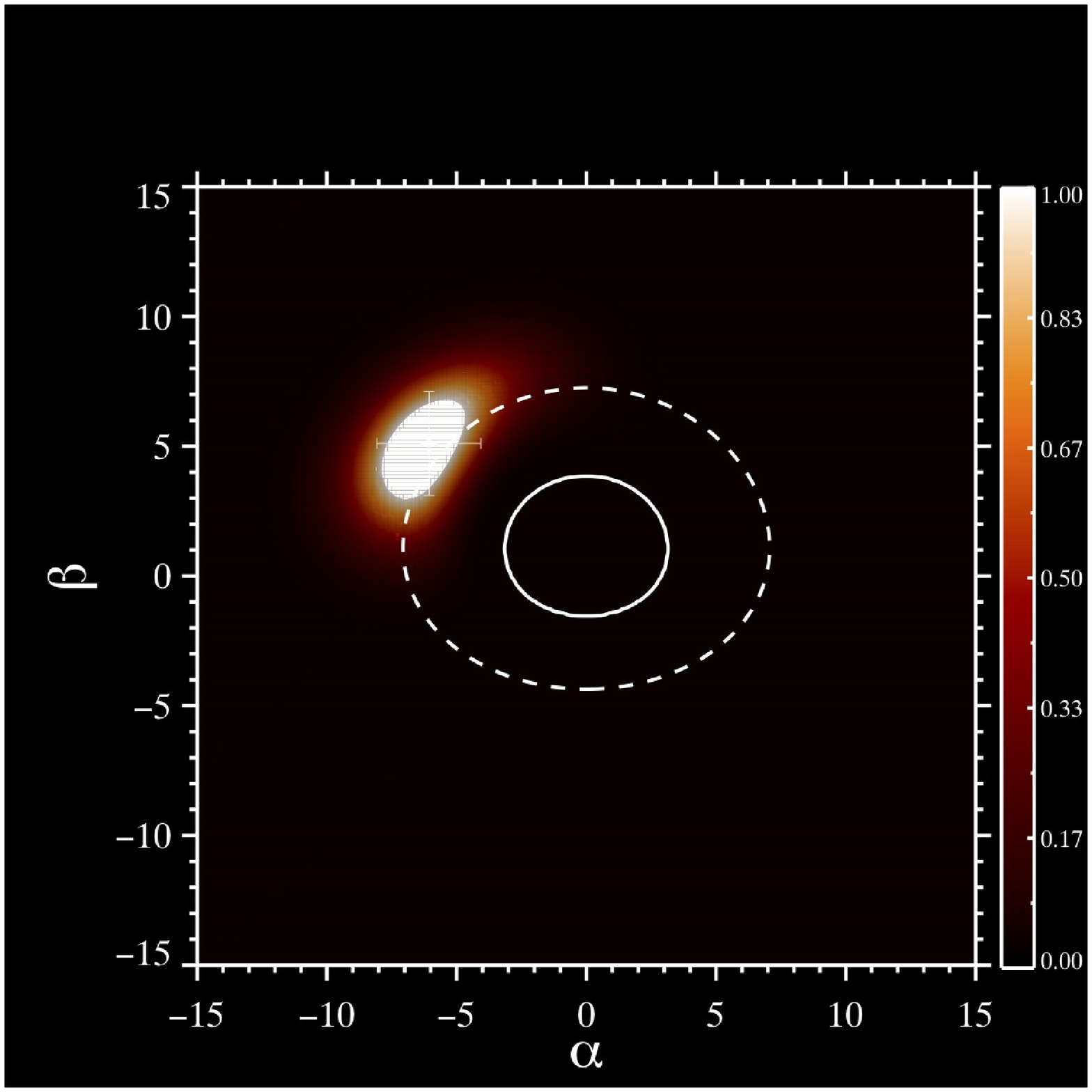}
\includegraphics[width=0.32\textwidth]{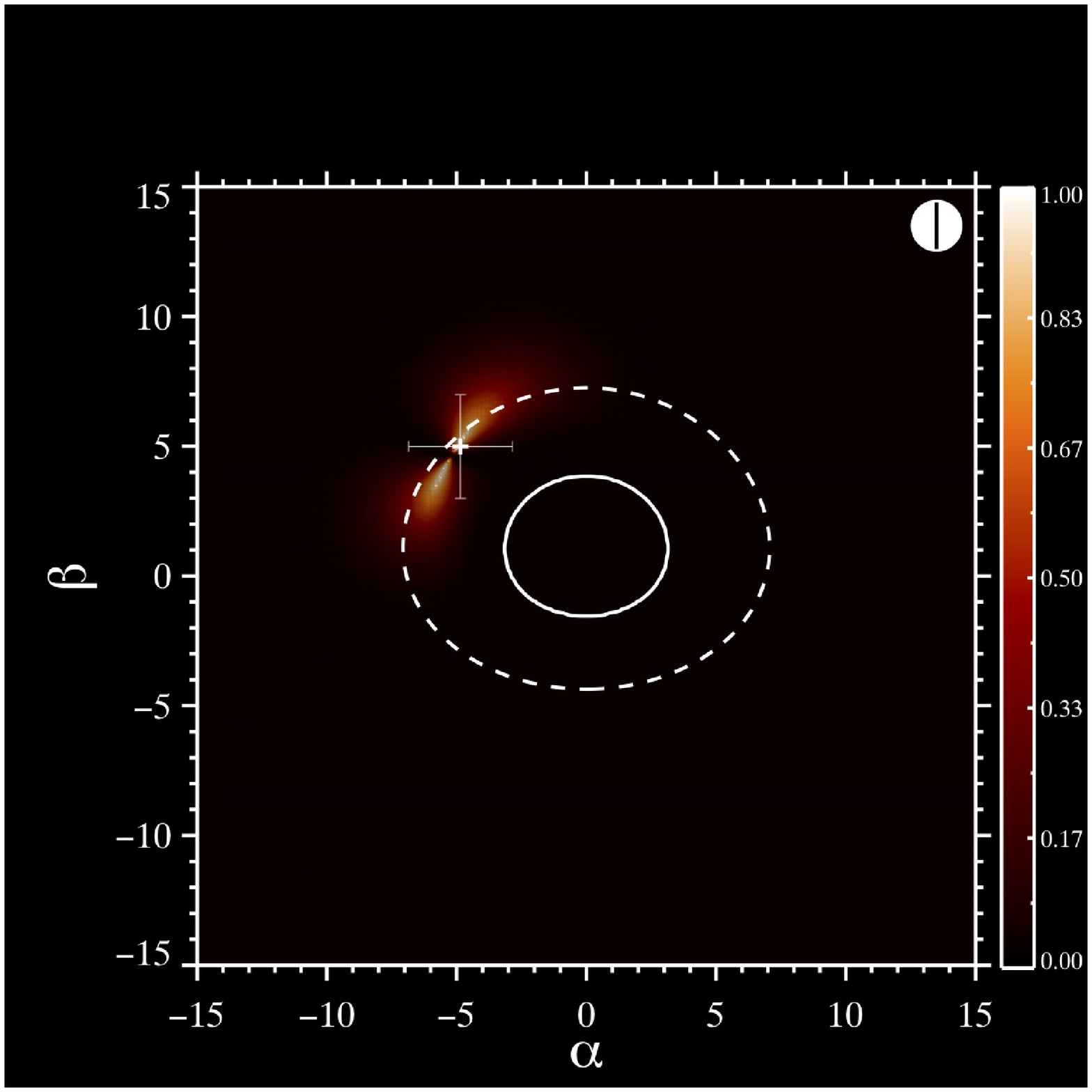} 
\includegraphics[width=0.32\textwidth]{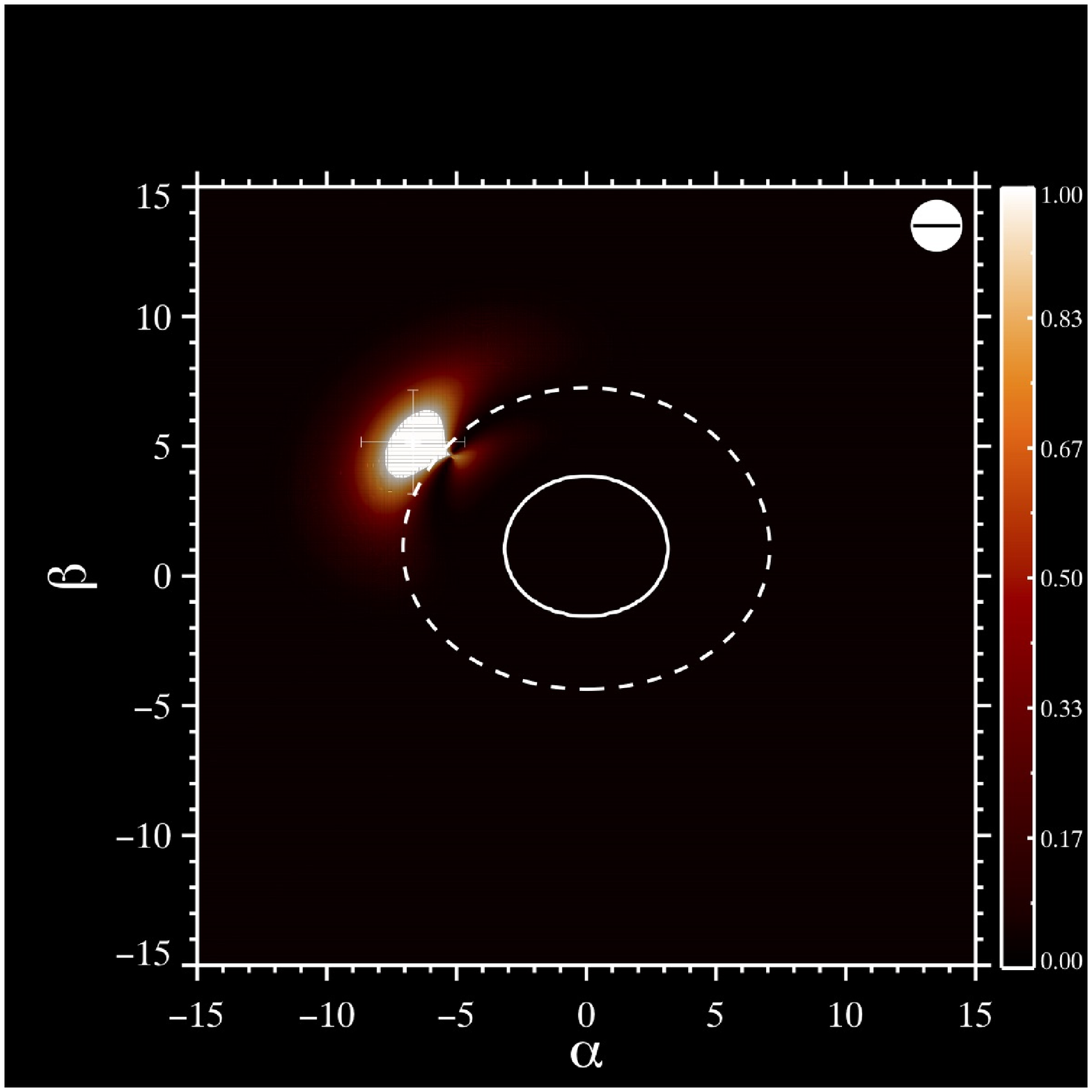}
\end{center}
\label{fig0}
\vspace*{1em}
\caption{An exemplary snapshot of a spot orbiting at constant radius just 
above the marginally stable orbit ($r=1.1r_{_{\rm ISCO}}$).
Left panel: the total intensity image shows the observer plane ($\alpha,\beta$) near a
non-rotating black hole, as observed at a moderate view angle 
($\theta_{\rm{}o}=45$~deg). The horizon radius (solid curve) and 
the ISCO (dashed curve) are indicated. Middle and right panels:
The spot
emission is assumed to be intrinsically polarized and recorded in
two polarization channels, rotated by 90 degrees with respect to each
other.\cite{zam10,zam11}}
\end{figure}

The geometrical effects of strong gravitational fields act on
photons independently of their energy.
The gravitational field is described by the metric of Kerr black hole 
\cite{Mis-Tho-Whe:1973}
\begin{equation}
ds^2=-\frac{\Delta}{\Sigma}\Big(dt-a\sin^2\theta\;d\phi\Big)^2
+\frac{\Sigma}{\Delta}\;dr^2+\Sigma\;d\theta^2
+\frac{\sin^2\theta}{\Sigma}\Big[a\;dt-\big(r^2+a^2\big)\;d\phi\Big]^2
\label{eq:metric}
\end{equation}
in Boyer-Lindquist (spheroidal) coordinates $t$, $r$, $\theta$, $\phi$. 
The metric functions $\Delta(r)$
and $\Sigma(r,\theta)$ are known in an explicit form. The event horizon occurs at 
the roots of equation $\Delta(r)=0$; the 
outer solution is found given by $r=R_+=1+(1-a^2)^{1/2}$. 

Let us note that there are some similarities as well as differences between
the expected manifestation of relativistic effects in polarization changes in X-rays and in
other spectral bands (NIR). We show these
interrelations in the associated poster and point out that the near-infrared polarization
measurements of the radiation flares from the immediate vicinity of the
horizon have been studied in detail from the Galactic Center (Sagittarius A*) 
supermassive black hole.\cite{mey06,zam10,zam11}

\section{Polarization from an Orbiting Spot}
Within the scheme of the spot scenario
the spots are considered to represent regions of enhanced emission on 
the disc surface rather than massive clumps that would
decay due to shearing motion in the disc. The observed signal is
modulated by relativistic effects. Doppler and
gravitational lensing influence the observed radiation flux and this can
be computed by GR ray-tracing. Such an approach has been developed
to compute also strong gravity effects acting on polarization properties.\cite{kar18}

\begin{figure}[tbh!]
\begin{center}
\includegraphics[width=0.32\textwidth]{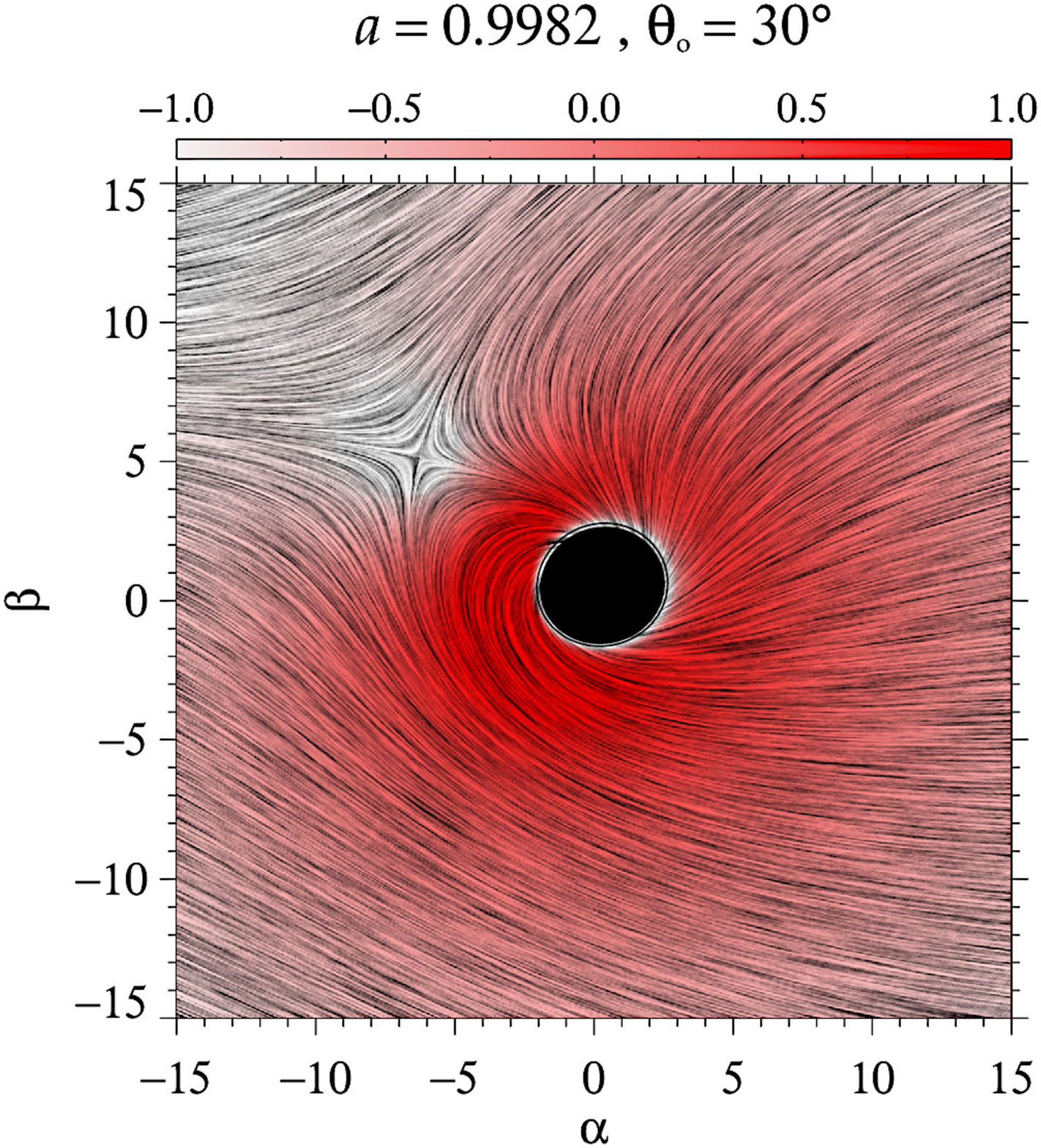}
\includegraphics[width=0.32\textwidth]{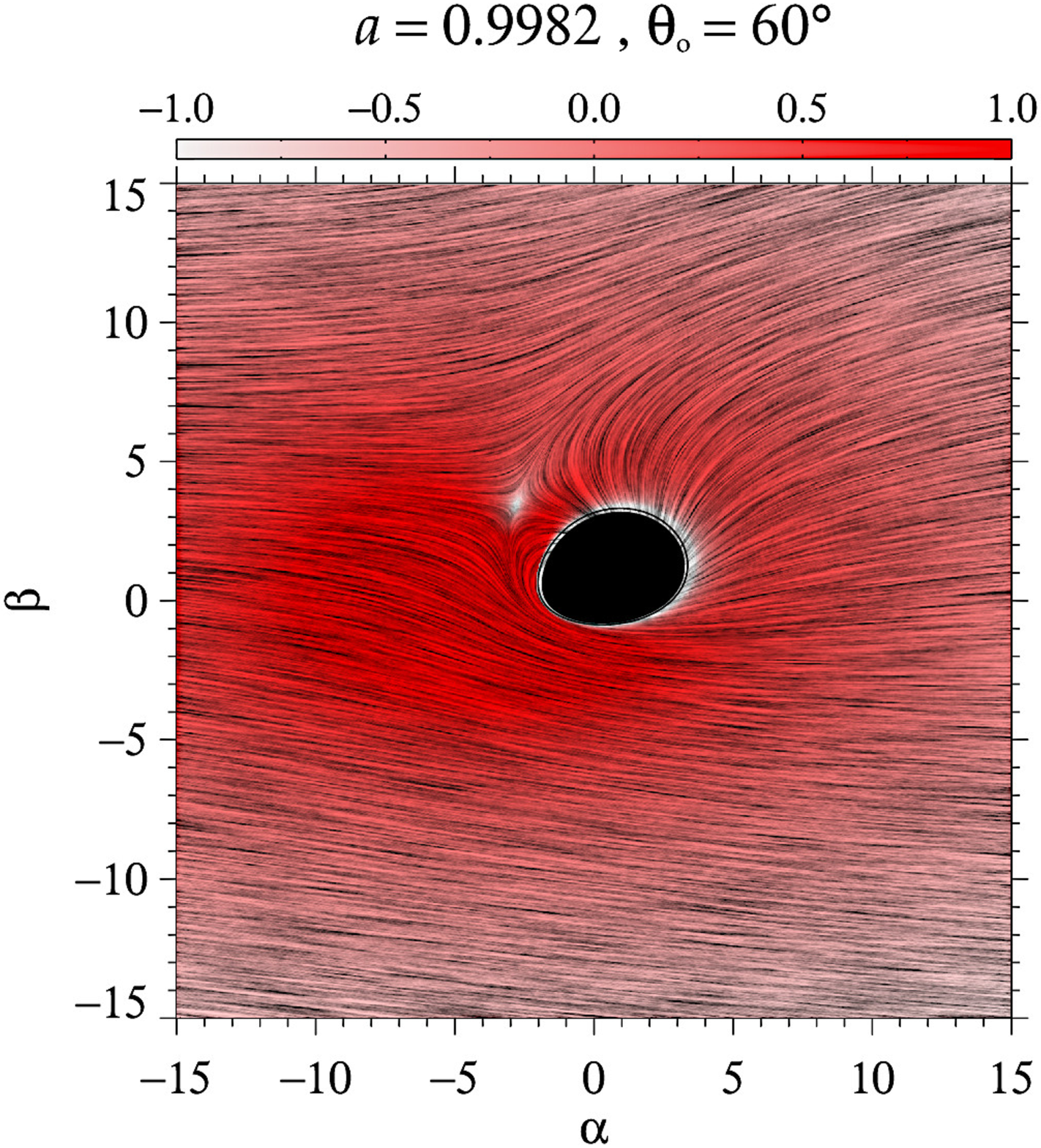}
\includegraphics[width=0.32\textwidth]{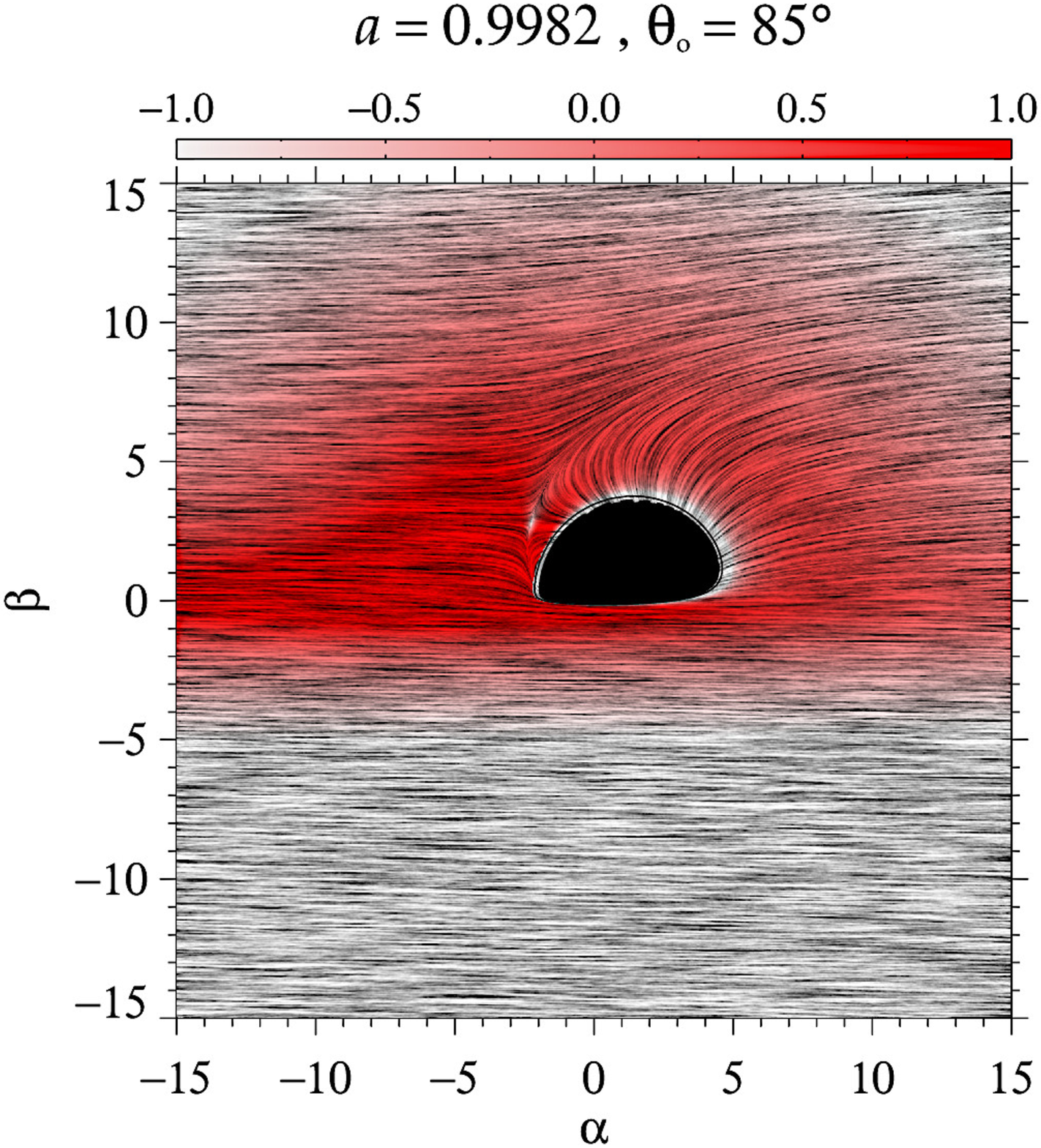}
\end{center}
\label{fig1}
\vspace*{1em}
\caption{A map of the distribution of the polarization angle (flow lines) and the intensity of the polarized light (colour coded in the logarithmic scale; arbitrary units are indicated by the colour bar), for a large value of spin $a$ of the Kerr black hole and different view angles of the observer, as indicated on top of the panels. This time the model has been computed for the case of thermal radiation of an equatorial accretion disk polarized via Comptonization; see Dov\v{c}iak et al.\ (2008).\cite{dov08}. A critical point is clearly seen in the flow-line structure through which then the orbiting spot circles.}
\end{figure}

Our code (KY) is publicly available.\citep{dov04} The current version\footnote{http://stronggravity.eu/results/models-and-data/} allows
the user to include the polarimetric resolution and to compute the observational
consequences of strong-gravity effects from a Kerr black hole accretion disc.
Within the XSPEC notation, this polarimetric resolution is encoded by a switch 
defining which of the four Stokes parameters is returned in the photon count array
at the moment of the output from the model evaluation. This way one can test 
and combine various models, and pass the resulting signal through the response 
matrices of different instruments. As an illustration,
figure~\ref{fig0} shows a single time-frame from a simulation of a spot rotating 
rigidly at constant radius.

The detected polarization degree is expected to decrease mainly in the part of the orbit where the spot moves
close to the region where the photons are emitted perpendicularly to the
disc and the polarization angle changes rapidly. The drop of the observed 
polarization degree occurs also in those parts
of the orbit where the magnetic field points approximately along light
ray (see fig.\ \ref{fig1} for a map showing the distribution of the polarization effects within
the disk plane of a rapidly rotating Kerr black hole). For the more realistic models the resulting polarization exhibits
a very complex behaviour due to, e.g., Faraday rotation effects interfering with
strong-gravity.\cite{dex18} Among generic features is the peak
in polarization degree for the spin parameter $a\rightarrow M$ and large
inclinations, caused by the lensing effect at a particular position of
the spot in the orbit. This effect disappears in the non-rotating $a=0$ case.

To conclude, let us note that the above described modelling has recently gained new
impetus in the context of forthcoming X-ray polarimetry satellite missions that have been 
under active consideration. The enhanced X-ray Timing and Polarimetry (eXTP) mission
\cite{zha16a,fer18} is the concept of novel Chinese X-ray mission that has currently reached the extended phase A
with a possible contribution from Europe and the anticipated launch date by 2025. 

The Large Area Detector (LAD) aboard 
the new mission will be a suitable instrument for performing time-resolved spectroscopy
with polarimetric resolution, large effective area, and moderate energy resolution to explore
bright flares from accreting black holes, possibly in the connection with the processes of tidal
disruption of stars and similarly violent events.\cite{zha16,zha19,zha19a,der19}

Let us just remind the reader that the revealing role of X-ray polarimetry clearly extends to 
all scales far away from the central black hole. This has been widely discussed in the context of 
reflection nebulae surrounding Galactic center.\cite{chu02,mar14} The combination of the polarization signal 
from the Sgr B and Sgr C complexes strengthens the need for an imaging detector with a 
fine spatial resolution to resolve the structures as small as the Bridge clouds. Therefore, 
the X-ray polarimetry is needed to explore variety of physical processes operating
near supermassive black hole.

\section{Conclusions}
We explored the approach based on mapping the Kerr black hole equatorial plane onto 
the observer's plane at radial infinity. Orbiting spots are projected onto the disc plane (hence
imposing the vertically averaged approximation) and then their image is transported
towards a distant observer. The strong gravity effects can be seen as the
predicted (time-dependent) direction of polarization is changed by light propagation
through the curved spacetime. What can be foreseen in the near future is the tracking of the
wobbling image centroid that a spot produces. With the polarimetric resolution, 
this wobbling can provide the evidence of orbiting features. 

In conclusion, the rotating spots are a viable scenario capable to explain the occurrence about 
once per day of modulated flares from within a few milli-arcseconds of the Sagittarius A* supermassive 
black hole.\cite{kar17,sha18} However, the astrophysically realistic scenarios have to account not
only for the time-scales of gradual dispersion of the orbiting features by tidal effects \cite{abr92,zak94} but also for the effects of plasma
influencing the light propagation through a non-vacuum spacetime near the black hole.\cite{bic75,kim19}

\section*{Acknowledgments}
The authors acknowledge the Czech Ministry of Education, Youth and Sports (M\v{S}MT) project INTER-INFORM (ref.\ LTI 17018),
 titled ``Promotion and Development of International Scientific Cooperation in Relativistic Astrophysics and Preparation of X-ray Space Missions''.

\end{document}